\documentclass[11pt,a4wide]{article}
\usepackage{multirow}
\usepackage{newlfont}
\usepackage{amsmath}
\usepackage{graphics}
\usepackage{float}
\usepackage{graphicx}
\usepackage{epsfig}
\usepackage[font=small,labelfont=bf,labelsep=space]{caption}
\begin{document}

\title{\emph{\textbf{Decay Properties of Conventional and Hybrid $B_c$ Mesons }}}
\author{\textbf{Nosheen Akbar}\thanks{E-mail: nosheenakbar@ciitlahore.edu.pk,noshinakbar@yahoo.com} \\
\textit{Department of Physics, COMSATS University Islamabad, Lahore Campus, } \\
{\textit{Defence Road, Lahore (54000), Pakistan}}}
\date{}
\maketitle

$\textrm{\textbf{Abstract}}$
--Spectrum, radial wave functions at origin, decay constants, weak decay widths, life time and branching ratios for radially excited conventional and hybrid $B_c$ mesons are derived within non-relativistic quark model framework employing Schr$\ddot{\textrm{o}}$dinger equation by shooting method. Calculated results are compared with available theoretical results and experimental observations. This work may help in identifying the new discovered $B_c$ meson states at CDF ,LHCb and ATLAS.

 \subsection*{1. INTRODUCTION}

\quad \quad $B_c$ meson consists of different flavored heavy quark--antiquark pair. It cannot be annihilated into gluons and can decay through weak interactions \cite{CDF} \cite{lodhi} \cite{0210381} only. Due to these characteristics, $B_c$ meson is considered a stable meson with narrow widths \cite{0406228}.

 $B_c$ meson was discovered in 1998 at Fermilab with mass $6.2749 \pm 0.0008$ GeV \cite{CDF}. After decades of ground state $B_c$ meson discovery, ATLAS Collaboration observed a peak for mass of excited state $B_c$ meson at $6.842 \pm 0.009 \textrm{GeV}$\cite{ATLAS}; however but LHCb Collaboration \cite{LHC} observed different results. Recently, CMS Collaboration \cite{CMS} observed two peaks for the excited states of $B_c$ meson, $B^+_c(2 ^1 S_0)$ and $B^{+\ast}_c(2 ^3 S_1)$. The mass of $B^+_c(2S)$ is measured to be $6.871 \pm 0.0012 \pm 0.0008 \pm 0.0080$ GeV \cite{CMS}. A mass difference of $0.0291 \pm 0.0015 \pm 0.0007$ GeV is measured between two states \cite{CMS}; however exact mass of $B^{+\ast}_c(2S)$ is still unknown.

Theoretically, a variety of techniques are available in the literature to study $B_c$ mesons like quark potential model \cite{nosheen19}-\cite{QiLe2019}, QCD sum rule \cite{1993}-\cite{9403208}, the heavy quark effective theory \cite{9412269}, lattice QCD \cite{0409090}-\cite{0305018} and  Dyson-Schwinger Equation \cite{2020} .

Meson under study ($B_c$) may be conventional or hybrid and may be identified through their masses and decay widths. Hybrid $B_c$ mesons are studied in ref. \cite{nosheen19}~\cite{1306.3486}. In this paper, non-relativistic potential model and its extended form is used to find masses and radial wave functions at origin for conventional as well as hybrid $B_c$ mesons using Born Openheimer formalism and adiabatic approximation. Decay constants, weak decay widths, life time and branching ratios for radial excited states of conventional and hybrid $B_c$ mesons are calculated using these masses and $\mid R(0) \mid^2$.

Potential models used for conventional and hybrid mesons are discussed in the Section 2 of this paper which were further used to calculate masses and radial wave functions at origin for the ground and radially excited state $B_c$ mesons by solving the Schr$\ddot{\textrm{o}}$dinger equation numerically. The expressions used to find decay constants, decay widths, life time and branching ratios of $B_c$ mesons are written in Section 3. The results are discussed in Section 4.

\section*{2. Potential Models for Conventional and Hybrid $B_c$ Mesons}

\quad \quad To study the properties of conventional $B_c$ mesons, following potential model\cite{0406228} is used:
\begin{equation}
 V(r) = \frac{-4\alpha _{s}}{3r} + b r + \frac{32\pi \alpha_s}{9 m_q m_{\overline{q}}} (\frac{\sigma}{\sqrt{\pi}})^3 e^{-\sigma ^{2}r^{2}} \textbf{S}_{q}. \textbf{S}_{\overline{q}} + \frac{4 \alpha _{s}}{m_q m_{\overline{q}} r^3} H_T + (\frac{\textbf{S}_{q}}{4 m_q^2}  + \frac{\textbf{S}_{\overline{q}}}{4 m_{\overline{q}}^2}).\textbf{L} (\frac{4\alpha _{s}}{3r^3}- \frac{b}{r})+ \frac{\textbf{S}_{q}+ \textbf{S}_{\overline{q}}}{2 m_q m_{\overline{q}}}.{\textbf{L}}\frac{4\alpha _{s}}{3r^3}. \label{Vr}
\end{equation}
Here, $\alpha _{s}$, $b$, $m_q$, $m_{\overline{q}}$ and $H_T$ are the strong coupling constants, string tension, mass of quark, mass of anti-quark and the tensor operator respectively. Here $m_q$ and $m_{\overline{q}}$ are the masses of $b$ and $c$ quarks, so can be written as $m_b$ and $m_c$.  Tensor operator is defined as:
\begin{equation}
H_T=\textbf{S}_q.{\hat{r}}\textbf{S}_{\overline{q}}.{\hat{r}}-\frac{1}{3}\textbf{S}_{q}.
\textbf{S}_{\overline{q}},
\end{equation}
such that
\begin{equation}
<^{3}L_{J}\mid H_T\mid ^{3}L_{J}>=\Bigg \{
\begin{array}{c}
-\frac{1}{6(2L+3)},J=L+1 \\
+\frac{1}{6},J=L \quad \quad \quad \quad \quad,\\
-\frac{L+1}{6(2L-1)},J=L-1
\end{array}
\end{equation}

Here, $L$ is the relative orbital angular momentum of the quark--antiquark and $S$ is the total spin angular momentum. In this paper, properties of $B_c$ mesons are study for pseudo scalar mesons (i.e. $L=0$). For $L=0$, spin orbit and tensor terms are equal to zero~\cite{charmonia05}. \\

\quad \quad To find the mass of $B_c$ mesons, numerical solution of the radial Schr$\ddot{\text{o}}$dinger equation
\begin{equation}
U^{\prime \prime }(r)+2\mu (E-V(r)-\frac{L(L+1)}{2\mu r^{2}})U(r)=0.
\label{P23}
\end{equation}
is found by employing the shooting method. Here $U(r)=rR(r)$, product of interquark distance $r$ and the radial wave function $R(r)$. At small distance (\textit{r} $\rightarrow$ 0), wave function becomes unstable due to very strong attractive potential. This problem is solved by applying smearing of position coordinates by using the method discussed in \cite{godfrey}. Parameters ($ m_{\overline{q}} = m_b = 4.733$ GeV, $ m_q = m_c = 1.3841$ GeV, $\alpha_s = 0.4035$, $b=0.18$, $\sigma = 1.0765$ GeV) are obtained by fitting the meson masses with corresponding experimentally known masses. Mass of the $\overline{b} c$ state is obtained after the addition of constituent quark masses in the energy $E$. Mass of pseudo-scalar and pseudo-vector$B_c$ meson states are reported in Table 1.

To study hybrid mesons, potential model for conventional mesons (mentioned above) is extended  by considering an additional term ($\frac{c}{r} + A\times \exp^{-Br^{0.3723}}$) \cite{Nosheen11}. Parameters $A = 3.4693 GeV$, $B  =1.0110 GeV$, and $c = 0.1745$ are taken from our earlier fit\cite{Nosheen11} to the lattice data \cite{Kuti97} of the parameters of the effective potential form corresponding to the first excited gluonic state. For hybrid mesons, radial Schrodinger equation is written as:
\begin{equation}
U^{\prime \prime }(r)+2\mu \left( E-V(r)-\frac{c}{r}-A\times \exp^{-Br^{0.3723}}-\frac{L(L+1)-2\Lambda
^{2}+\left\langle J_{g}^{2}\right\rangle}{2\mu r^{2}}\right) U(r)=0,
\end{equation}
Here, $J_{g}$ is the gluon angular momentum and $\Lambda$ is the projection of gluon angular momentum. For the first gluonic excitation, $\left\langle J_{g}^{2}\right\rangle =2$ and $\Lambda =1$~\cite{Kuti97}. The masses of hybrid $B_c$ meson states are found by solving the above equation by shooting method and results are reported in Table 2.

\section*{3. \textit{Decay properties of conventional and hybrid $B_c$ Mesons}}

\subsection*{3.1. Decay Constants}
 Decay constant is an important characteristic of mesons. Decay constants ($f_p$) of pseudo scalar and pseudo vector mesons depend on $|R(0)|^2 $. Following Van-Royen-Weisskopf formula \cite{VWF} is used to find decay constants.
\begin{equation}
f_p = \sqrt{\frac{3 |R(0)^2|}{\pi M_p}} =\sqrt{\frac{12 |\psi(0)^2|}{M_p}}.
\end{equation}
where $M_p$ is the mass of the corresponding $B_c$ meson. By incorporating the first order QCD correction factor, the decay constant can be written as:
\begin{equation}
f_p = \sqrt{\frac{3 |R(0)^2|}{\pi M_p}} (1-\frac{\alpha_s}{\pi} \big[ \triangle - \frac{m_Q - m_{\overline{Q}}}{m_Q - m_{\overline{Q}}} ln\mid\frac{m_Q}{m_{\overline{Q}}}\mid \big]),
\end{equation}
where $\triangle=2$ for $^1 S_0$ mesons and $\triangle=8/3$ for $^3 S_1$ mesons.
Decay constants for conventional and hybrid mesons are reported in Tables(3-4).

\subsection*{3.2. Weak Decay}

According to the spectator model \cite{lodhi}, $B_c$ meson decay can be divided into three classes
(i) $b$-quark decay having $c$-quark as spectator,
(ii) $c$-quark decay having $b$-quark as spectator,
(iii) annihilation decay of $B_c$ meson.
$b$-quark and $c$-quark decay widths of $B_c$ meson can be calculated by following expressions ~\cite{lodhi}\cite{0406228}\cite{2012}:
\begin{equation}
\Gamma(b \rightarrow X) = \frac{9 G^2_F \mid V_{cb}\mid ^2 m^5_b}{192 \pi^3},
\end{equation}
\begin{equation}
\Gamma(c \rightarrow X) = \frac{5 G^2_F \mid V_{cs}\mid ^2 m^5_c}{192 \pi^3},
\end{equation}
 where $\mid V_{cb}\mid$ and $\mid V_{cs}\mid$ are the $cb$ and $cs$ elements of CKM matrix. I used $\mid V_{cs} \mid =0.9736$ and $\mid V_{cb} \mid = 0.04214$ taken from PDG \cite{pdg}. $G_F$ is the fermi constant and equal to $1.166 \times 10^{-5}$ $\textrm{GeV}^{2}$. With this data, i calculated the partial decay widths as
\begin{equation*}
\Gamma(b \rightarrow X) = 8.669 \times 10^{-4} \textrm{eV},\quad \quad \Gamma(c \rightarrow X) = 5.497 \times 10^{-4} \textrm{eV}.
\end{equation*}

Annihilation decay for pseudoscalar $B_c$ meson states (conventional and hybrid) can be calculated by the following relation defined in ~\cite{lodhi} \cite{0406228}\cite{2012}:
\begin{equation}
\Gamma(anni) = \frac{G^2_F \mid V_{cb}\mid ^2 f^2_{bc} M_{B_c} m^2_q(1-\frac{m^2_q}{M^2_{B_c}})C_q}{8 \pi},
\end{equation}
where $C_q = 3 \mid V_{cs}\mid ^2$ for $c \overline{s}, u \overline{s}$ and $C_q = 1$ for $ \ell \upsilon_{\ell}$ where $\ell=e,\mu, \tau$. In the present paper, annihilation decay is calculated for $ \tau \upsilon_{\tau}$ and $c \overline{s}$.

Total decay width can be roughly estimated as the sum of $b$-quark, $c$-quark and annihilation decay,i.e;
\begin{equation}
\Gamma(B_c \rightarrow X) = \Gamma(b \rightarrow X)+ \Gamma(c \rightarrow X) + \Gamma(anni).
\end{equation}
For conventional mesons, decay widths for Pseudo scalar $B_c$ mesons are reported in Table 5, while Table 7 contains the decay widths for hybrid mesons.
\subsection*{3.3. Life time}
Life-time($\tau$) of conventional and hybrid $B_c$ mesons is found by following relation:
\begin{equation}
\tau = \frac{\hbar}{\Gamma}
\end{equation}
where $\hbar = 6.582 \times 10^{-25}$ $\textrm{GeV.sec}$

\subsection*{3.4. Branching ratio}
Branching ratios of $b$-decay, $c$-decay and annihilation decay are found by taking ratio of corresponding decay width to total decay width. i.e,
\begin{equation}
\begin{array}{c}
{\ss}(b \rightarrow X)=\frac{\Gamma(b \rightarrow X)}{\Gamma(B_c \rightarrow X)}, \\
\\
{\ss}(c \rightarrow X)=\frac{\Gamma(c \rightarrow X)}{\Gamma(B_c \rightarrow X)},\\
\\
{\ss}(anni.)=\frac{\Gamma(anni.)}{\Gamma(B_c \rightarrow X)}.
\end{array}
\end{equation}
\section*{4. DISCUSSION AND CONCLUSION}

\quad \quad Based on nonrelativistic potential model, masses are calculated for the ground as well as radially excited states of conventional and hybrid $B_c$ mesons and reported in Tables (1-2) along with the experimental and theoretical predictions of other works. The results reported in Table 3 illustrate that radial wave functions at origin and decay constants for conventional mesons are decreasing toward higher radial excitations. But the situation is reversed in case of hybrids. For hybrids, radial wave functions at origin are increasing toward higher radial excitations. Pseudoscalar $B_c$ mesons have higher values of $|R(0)|^2$ and $f_p$ as compared to vector mesons. It is observed that hybrids have lesser value of decay constant than the conventional mesons. Decay width, life time and branching ratio of conventional and hybrid $B_c$ mesons are reported in Tables (5-7). In Table 6, conventional $1 ^1S_0$ state decay width, life time and branching ratio are compared with others. It is observed that my calculated life time is more close to the experimental result\cite{pdg}  as compare to others theoretical works. Branching ratio of $b$-decay, $c$- decay and annihilation decay for conventional meson are also very close to others theoretical results \cite{0406228}. For conventional $B_c$ mesons, branching ratios of annihilation decays are very small as compare to $b$-decay and $c$-decay. In case of hybrid $B_c$ mesons branching ratios for annihilation decays are less than 1. This show that annihilation decay is rare for $B_c$ mesons. The agreement of results with others work shows the validity of my method. Using the extended potential model for hybrids, decay widths, life time and branching ratios are predicted for the hybrid $B_c$ mesons. It is observed that hybrids have less decay widths as compare to conventional $B_c$ meson. From these results, i conclude that hybrid $B_c$ mesons are more stable than conventional mesons with the same $L$ and $S$. This work may help in recognizing the higher states of $B_c$ mesons that will discover at CDF detector, LHCb, and ATLAS.

\section*{Acknowledgement}
\quad \quad NA acknowledge the financial support of Higher Education Commission of Pakistan through NRPU project 7969.

\begin{table}\caption{Masses of ground, radially, and orbitally excited state of $B_c$ mesons; calculated masses are rounded to 0.0001 GeV.}
\tabcolsep=4pt
\fontsize{7}{10}\selectfont
\begin{center}
\begin{tabular}{|c|c|c|c|c|c|c|c|}
\hline
Meson & $J^{P}$& My Calculated & Theor. Mass & Lattice & Theor. Mass(GI) &  Exp. Mass \\
State & & Mass, \textrm{GeV}  & \cite{nosheen19}, \textrm{GeV} &\cite{lattice}& \cite{0406228}, \textrm{GeV} &\cite{pdg}, \textrm{GeV} \\ \hline
 $ (1 ^3S_1)$ & $1^{-}$ & 6.3062 & 6.314 & $6.331 \pm 0.004 \pm 0.006$ &6.338 & \\
$ (1 ^1S_0)$ & $0^{-}$ & 6.2740 & 6.274 & $6.276 \pm 0.003 \pm 0.006$ & 6.271 &$ 6.2749 \pm 0.0008$ \\ \hline
 $(2 ^3S_1)$& $1^{-}$ &6.8845 & 6.855 & &6.887&\\
$ (2 ^1S_0)$ & $0^{-}$ &6.8717 & 6.841 & &6.855& $ 6.871 \pm 0.0012 \pm 0.0008 \pm 0.0008 $\\ \hline
 $(3 ^3S_1)$& $1^{-}$ &7.2905 & 7.206 & & 7.272 &\\
 $(3 ^1S_0)$ & $0^{-}$ &7.2818 & 7.197 & &7.250 &\\ \hline
 $(4 ^3S_1)$& $1^{-}$ & 7.6323& 7.495& & &\\
 $(4 ^1S_0)$ & $0^{-}$ & 7.6255 &7.488 & & & \\ \hline
$(5 ^3S_1)$& $1^{-}$ & 7.9374& - & & &\\
 $(5 ^1S_0)$ & $0^{-}$ & 7.9317 & - & & &\\ \hline
 $(6 ^3S_1)$& $1^{-}$ & 8.2178& - &  & &\\
 $(6 ^1S_0)$ & $0^{-}$ & 8.2129 & - & & &\\ \hline

\end{tabular}
\end{center}
\end{table}

\begin{table}\caption{Masses of ground, radially, and orbitally excited state of $B_c$ hybrid mesons;}
\tabcolsep=4pt
\fontsize{7}{10}\selectfont
\begin{center}
\begin{tabular}{|c|c|c|c|} \hline
Meson & $J^{P}$& My Calculated & Mass  \\
State & & Mass, \textrm{GeV} & \cite{nosheen19},\textrm{GeV} \\ \hline
 $ (1 ^3S_1)$ & $1^{-}$ & 7.410 & 7.422 \\
$ (1 ^1S_0)$ & $0^{-}$ & 7.4026 & 7.415\\ \hline
 $(2 ^3S_1)$& $1^{-}$ & 7.7081 &7.654 \\
$ (2 ^1S_0)$ & $0^{-}$ & 7.6999 &7.646\\ \hline
 $(3 ^3S_1)$& $1^{-}$ & 7.9854 & 7.874\\
 $(3 ^1S_0)$ & $0^{-}$ & 7.9779 & 7.866\\ \hline
 $(4 ^3S_1)$& $1^{-}$ &8.246 & 8.082\\
 $(4 ^1S_0)$ & $0^{-}$ & 8.2392 & 8.075\\ \hline
$(5 ^3S_1)$& $1^{-}$ &8.4929 & \\
 $(5 ^1S_0)$ & $0^{-}$ & 8.4869 &\\ \hline
 $(6 ^3S_1)$& $1^{-}$ & 8.7287 &\\
 $(6 ^1S_0)$ & $0^{-}$ & 8.7233 &\\ \hline

\end{tabular}
\end{center}
\end{table}

\begin{table}\caption{Radial wave function at origin and decay constant of $B_c$ mesons . }
\tabcolsep=4pt
\fontsize{8}{10}\selectfont
\begin{center}
\begin{tabular}{|c|c|c|c|c|c|c|c|c|c|c|}
\hline
Meson & $J^{P}$& My Calculated & $|R(0)|^2$ & without correction & with correction &$f_p$ &$f_p$ &$f_p$ & $f_p$ & $f_p$ \\
State & & $|R(0)|^2$, $\textrm{GeV}^3$ & \cite{054025}, $\textrm{GeV}^3$ & $f_p$, $\textrm{GeV}$ & $f_p$, $\textrm{GeV}$ & \cite{054025}, $\textrm{GeV}$  &\cite{13100941}, $\textrm{GeV}$ &\cite{a}, $\textrm{GeV}$ &\cite{2001}, $\textrm{GeV}$& \cite{soni},$\textrm{GeV}$\\ \hline
  $ (1 ^3S_1)$ & $1^{-}$ &1.9566 & -& 0.5443 & 0.4049 & 0.471&- &0.411 & 0.604 &0.435\\
$ (1 ^1S_0)$ & $0^{-}$ & 2.2393 &1.994 &0.5839 & 0.4844 & 0.498 & $0.528 \pm 0.019 $&0.412 & 0.607 &0.432\\ \hline
 $(2 ^3S_1)$& $1^{-}$ &1.2018 &- & 0.4083 & 0.3038 & && & &0.356\\
$ (2 ^1S_0)$ & $0^{-}$ &1.2605 &1.144 &0.4185 &0.3472& & & & &0.355\\ \hline
 $(3 ^3S_1)$& $1^{-}$ &1.0083 & -& 0.3634 &0.2703& & & & &0.326\\
 $(3 ^1S_0)$ & $0^{-}$  & 1.0399 & 0.944&0.3693 & 0.3064& & & & &0.325\\ \hline
 $(4 ^3S_1)$& $1^{-}$ &0.9138 & -& 0.3381 &0.2515 & & & & &0.308\\
 $(4 ^1S_0)$ & $0^{-}$ & 0.9358& 0.8504 & 0.3423 &0.284 & & & & &0.307\\ \hline
 $(5 ^3S_1)$& $1^{-}$ &0.8553 & -& 0.3208 &0.2387 & & & & &0.308\\
 $(5 ^1S_0)$ & $0^{-}$ & 0.8726& - & 0.3423 & 0.2688 & & & & &0.307\\ \hline
$(6^3S_1)$& $1^{-}$ &0.8148 & -& 0.3077 & 0.2289 & & & & &0.308\\
 $(6 ^1S_0)$ & $0^{-}$ & 0.8293 & - & 0.3105 & 0.2576 & & & & &0.307\\ \hline
\end{tabular}
\end{center}
\end{table}

\begin{table}\caption{Radial wave function at origin and decay constant of hybrid $B_c$ mesons . }
\tabcolsep=4pt
\fontsize{8}{10}\selectfont
\begin{center}
\begin{tabular}{|c|c|c|c|c|c|}
\hline
Meson & $J^{P}$& My Calculated & without correction & with correction  \\
State & & $|R(0)|^2$, $\textrm{GeV}^3$ & $f_p$, $\textrm{GeV}$ & $f_p$, $\textrm{GeV}$\\ \hline
  $ (1 ^3S_1)$ & $1^{-}$ &0.2061 & 0.163 & 0.1213 \\
$ (1 ^1S_0)$ & $0^{-}$ & 0.2533  &0.1807 & 0.1499 \\ \hline
 $(2 ^3S_1)$& $1^{-}$ &0.2942 & 0.191 & 0.142 \\
$ (2 ^1S_0)$ & $0^{-}$ &0.3426 &0.206 &0.1709 \\ \hline
 $(3 ^3S_1)$& $1^{-}$ & 0.3391&  0.2015 &0.1499 \\
 $(3 ^1S_0)$ & $0^{-}$  & 0.3797 &0.2131 & 0.1768 \\ \hline
 $(4 ^3S_1)$& $1^{-}$ &0.3658&  0.2059 &0.1532 \\
 $(4 ^1S_0)$ & $0^{-}$ & 0.3991& 0.215 &0.1783 \\ \hline
 $(5 ^3S_1)$& $1^{-}$ &0.3834 &0.2077 &0.1545 \\
 $(5 ^1S_0)$ & $0^{-}$ & 0.4111&  0.215 & 0.1784 \\ \hline
$(6^3S_1)$& $1^{-}$ &0.3972 & 0.2085 & 0.1551 \\
 $(6 ^1S_0)$ & $0^{-}$ & 0.4206 & 0.2145 &0.178 \\ \hline
\end{tabular}
\end{center}
\end{table}

\begin{table}\caption{Annihilation decay, total decay width, life-time and branching ratio of pseudo scalar $B_c$ mesons . }
\tabcolsep=4pt
\fontsize{8}{10}\selectfont
\begin{center}
\begin{tabular}{|c|c|c|c|c|c|c|c|}
\hline
& & Annihilation & Total decay width & life time & ${\ss}(b\rightarrow X)$ & ${\ss}(c\rightarrow X)$ & ${\ss}(anni.)$ \\
& State & $10^{-12}$ ,$\textrm{GeV}$ & $10^{-12}$, $\textrm{GeV}$ &  $\textrm{sec}$ & $ \% $  & $ \% $&$ \% $ \\ \hline

\multirow{6}{*}{$c\overline{s}$}&$ (1 ^1S_0)$ & 0.0732 & 1.49 & 0.4418 & 58 &36.9 & 4.9  \\
&$ (2 ^1S_0)$ &  0.0416 & 1.458 & 0.4514 & 59.4 &37.7 & 2.8 \\
 &$(3 ^1S_0)$ &  0.0345 & 1.451 & 0.454 & 59.7 &37.9 & 2.4 \\
 &$(4 ^1S_0)$ &  0.0311 & 1.448 & 0.455 & 59.9 & 38 & 2.1 \\
 &$(5 ^1S_0)$ &  0.0291 & 1.446 & 0.455 & 60 &38 & 2 \\
 &$(6 ^1S_0)$ &  0.027 & 1.444 & 0.456 & 60 &38 & 1.9 \\ \hline \hline

\multirow{4}{*}{$\tau \nu_\tau$} & $ (1 ^1S_0)$ & 0.0411 & 1.4576 & 0.45 & 59.5 &37.7 & 2.8  \\
&$ (2 ^1S_0)$ &  0.0234 & 1.44 & 0.457 & 60 &38 & 1.6 \\
&$(3 ^1S_0)$ &  0.0195 & 1.436 & &  & & \\
 &$(4 ^1S_0)$ &  0.0176 & 1.434 & & & & \\ \hline
\end{tabular}
\end{center}
\end{table}

\begin{table}\caption{ Comparison of annihilation decay, total decay width, life-time and branching ratio of pseudo scalar $B_c$ mesons . }
\tabcolsep=4pt
\fontsize{8}{10}\selectfont
\begin{center}
\begin{tabular}{|c|c|c|c|c|c|c|c|c|}
\hline
& & & Annihilation & Total decay width & life time & ${\ss}(b\rightarrow X)$ & ${\ss}(c\rightarrow X)$ & ${\ss}(anni.)$ \\
& & State & $10^{-4}$ ,$\textrm{eV}$ & $10^{-4}$, $\textrm{eV}$ &  $\textrm{sec}$ & $ \% $  & $ \% $&$ \% $ \\ \hline

\multirow{6}{*}{$c\overline{s}$}& &$ (1 ^1S_0)$ & 0.732 & 14.9 & 0.4418 & 58 &36.9 & 4.9  \\
 &\cite{2012} &$ (1 ^1S_0)$ & 1.17 & 19.17 & 0.344 &  & &  \\
 &\cite{0406228} &$ (1 ^1S_0)$ & 0.67& 8.8 & 0.75 & 54 & 38 & 8  \\
 &\cite{lodhi} &$ (1 ^1S_0)$ & 1.4 & 14 & 0.47 &  & &  \\
&\cite{pdg} &$ (1 ^1S_0)$ &  & & 0.46 &  & &  \\
\multirow{1}{*}{$\tau \nu_\tau$} & & $ (1 ^1S_0)$ & 0.0411 & 1.4576 & 0.45 & 59.5 &37.7 & 2.8  \\ \hline
\end{tabular}
\end{center}
\end{table}

\begin{table}\caption{annihilation decay, total decay width , life-time and branching ratio of hybrid $B_c$ mesons . }
\tabcolsep=4pt
\fontsize{8}{10}\selectfont
\begin{center}
\begin{tabular}{|c|c|c|c|c|c|c|c|}
\hline
& & Annihilation & Total decay width & life time & ${\ss}(b\rightarrow X)$ & ${\ss}(c\rightarrow X)$ & ${\ss}(anni.)$ \\
& State & $10^{-12}$ ,$\textrm{GeV}$ & $10^{-12}$, $\textrm{GeV}$ &  $\textrm{sec}$ & $ \% $  & $ \% $&$ \% $ \\ \hline
\multirow{6}{*}{$c\overline{s}$}&$ (1 ^1S_0)$ & 0.0084 & 1.425 & 0.462 & 60.8 &38.6 & 0.6  \\
&$ (2 ^1S_0)$ &  0.014 & 1.428 & 0.461 & 60.7 &38.5 & 0.98 \\
 &$(3 ^1S_0)$ &  0.0126 & 1.429 & 0.46 & 60.7 &38.5 & 0.88 \\
 &$(4 ^1S_0)$ &  0.0133 & 1.43 & 0.46 & 60.6 & 38.4 & 0.93 \\
 &$(5 ^1S_0)$ &  0.0138 & 1.43 & 0.46 & 60.6 &38.4 & 0.96 \\
 &$(6 ^1S_0)$ &  0.014 & 1.431 & 0.46 & 60.6 &38.4 & 0.98 \\ \hline \hline

\multirow{4}{*}{$\tau \nu_\tau$}&$ (1 ^1S_0)$ & 0.0048 & 1.421 & 0.46 & 61 &38.7 & 0.3  \\
&$ (2 ^1S_0)$ & 0.0065 & 1.423 & 0.46 & 60.9 &38.6 & 0.45 \\
&$(3 ^1S_0)$ & 0.0072 & 1.424 & &  & & \\
 &$(4 ^1S_0)$ & 0.0076 & 1.424 & & & & \\ \hline
\end{tabular}
\end{center}
\end{table}


\begin{thebibliography}{5}
\bibitem{CDF}
F. Abe et al. (CDF Collab.), Phys. Rev. Lett. \textbf{81}, 2432 (1998); Phys. Rev. D $\textbf{58}$ 112004 (1998); P. Ball et al., CERN-TH/2000-101, hep-ph/0003238.

\bibitem{lodhi}
A Adb El-Hady, M A K Lodhi and J P Vary, Phys. Rev. D \textbf{59}, 094001 (1999).

\bibitem{0210381}
{\normalsize D. Ebert, R. N. Faustov, V. O. Galkin, Phys. Rev. D \textbf{67}, 014027 (2003). }

\bibitem{0406228}
{\normalsize S. Godfrey, Phys. Rev. D \textbf{70}, 054017 (2004). }

\bibitem{ATLAS}
G. Aad et al. (ATLAS Collab.), Phys. Rev. Lett. \textbf{113}, 212004 (2014).

\bibitem{LHC}
R. Aaij et al. (LHCb Collab.), JHEP \textbf{1801}, 138 (2018).

\bibitem{CMS}
A. M. Sirunyan et al. (CMS Collab.), Phys. Rev. Lett. \textbf{122}, 132001 (2019).

\bibitem{nosheen19}
N. Akbar, F. Akram, B. Masud, and M. A. Sultan, Eur. Phys. J. A \textbf{55}, 82 (2019).

\bibitem{054025}
E. J. Eichten and C. Quigg, Phys. Rev. D \textbf{99}, 054025 (2019).

\bibitem{ishrat19}
I. Asghar, F. Akram, B. Masud and M. A. Sultan, arXiv : 1910.02680v1.

\bibitem{9402210}
E. J. Eichten and C. Quigg, Phys. Rev. D \textbf{49}, 5845 (1994).

\bibitem{9511267}
{\normalsize S. N. Gupta and J. M. Johnson, Phys. Rev. D \textbf{53}, 312 (1996). }

\bibitem{9806444}
{\normalsize L. P. Fulcher, Phys. Rev. D \textbf{60}, 074006 (1999).}

\bibitem{054016}
A. P. Monteiro, M. Bhat, and K. B. Vijaya Kumar, Phys. Rev. D \textbf{95}, 054016 (2017).

\bibitem{1750021}
A. P. Monteiro, M. Bhat, and K. B. Vijaya Kumar, Int. J. Mod. Phys. A \textbf{32},
1750021 (2017).

\bibitem{2012}
K. K. Pathak and D K CHOUDHURY, PRAMANA journal of Physics, \textbf{79}, No. 6, 1385 (2012)

\bibitem{1504.07538}
{\normalsize M. Abu-Shady, Int. J. Appl. Math. Theor. Phys. \textbf{2}, 16 (2016). }

\bibitem{QiLe2019}
Q. Li, M.-S. Liu, L.-S. Lu, Q.-F. Lü, L.-C. Gui, and X.-H. Zhong, Phys. Rev. D \textbf{99}, 096020 (2019).

\bibitem{1993}
C. A. Dominguez, K. Schilcher Y. L. Wu, Physics Letters B \textbf{298}, 190  (1993).

\bibitem{9406339} {\normalsize S. S. Gershtein, V. V. Kiselev, A. K. Likhoded, and A. V. Tkabladze, Phys. Rev. D \textbf{51}, 3613 (1995). }

\bibitem{1306.3486}
{\normalsize W. Chen, T. G. Steele, and Shi-Lin Zhu, J. Phys. G: Nucl. Part. Phys. \textbf{41}, 025003 (2014). }

\bibitem{9403208}
{\normalsize E. Bagan, H. G. Dosch, P. Gosdzinsky, S. Narison, and J.-M. Richard, Z. Phys. C \textbf{64}, 57 (1994). }

\bibitem{9412269}
{\normalsize J. Zeng, J. W. Van Orden, and W. Roberts, Phys. Rev. D \textbf{53}, 5229 (1995). }

\bibitem{0409090}
{\normalsize I. F. Allison, C. T. H. Davies, A. Gray, A. S. Kronfeld, P. B. Mackenzie, J. N. Simone,(HPQCD Collab., FNAL lattice Collab., UKQCD Collab.) Nucl. Phys. Proc. Suppl. \textbf{140,} 440 (2005). }

\bibitem{Lattice}
{\normalsize C. T. H. Davies, K. Hornbostel, G. P. Lepage, A. J. Lidsey, J. Shigemitsu, and J. Sloan, Phys. Lett. B \textbf{382}, 131 (1996). }

\bibitem{0305018}
{\normalsize G. M. de Divitiis, M. Guagnelli, F. Palombi, R. Petronzio, and N. Tantalo,, Nucl. Phys. B \textbf{675}, 309 (2003). }

\bibitem{2020}
Muyang Chen, Lei Chang, Yu-xin Liu, Phys. Rev. D \textbf{101}, 056002 (2020).

\bibitem{charmonia05}
{\normalsize T. Barnes, S. Godfrey, and E. S. Swanson, Phys. Rev. D \textbf{72}, 054026 (2005). }

\bibitem{godfrey}
S. Godfrey and N. Isgur, Phys. Rev. D \textbf{32}, 189 (1985).

\bibitem{Nosheen11}
N. Akbar, B. Masud, S. Noor, Eur. Phys. J. A $textbf{47}$, 124 (2011); erratum: Eur. Phys. J. A \textbf{50}, 121 (2014).

\bibitem{Kuti97}
K. J. Juge, J. Kuti and C. J. Morningstar, Nucl. Phys. Proc. Suppl. \textbf{63}, 326 (1998).

\bibitem{VWF}
Van Royen R et al., Nuovo Cimento 50, (1967).

\bibitem{pdg}
M. Tanabashi et al. (Particle Data Group), Phys. Rev. D \textbf{98}, 030001 (2018).

\bibitem{13100941}
 M. J. Baker, J. Bordes, C. A. Dominguez, J. Pearrocha, and K. Schilcher, JHEP \textbf{1407}, 032 (2014).

\bibitem{lattice}
N. Mathur, M. Padmanath, and S. Mondal, Phys. Rev. Lett. \textbf{121}, 202002 (2018).

\bibitem{a}
N. R. Soni and J. N. Pandya, Proceedings of the DAE Symp. on Nucl. Phys. \textbf{58}, 674 (2013).

\bibitem{2001}
J. N. Pandya and P. C. Vinodkumar, Pramana J. Phys. \textbf{57}, 821 (2001).

\bibitem{soni}
N. R. Soni, B. R. Joshi, R. P. Shah, H. R. Chauhan, and J. N. Pandya, Eur. Phys. J. C 78, 592 (2018).

\bibitem{wong04}
C. -Y. Wong, Phys. Rev. C \textbf{69}, 055202 (2004).

\end{thebibliography}
\end{document}